\documentclass[reprint,twocolumn,showpacs,pas,prl,english]{revtex4}
\usepackage{graphicx}%
\usepackage{dcolumn}%
\usepackage{bm}%

\makeatletter



\usepackage{babel}
\makeatother

\begin{document}

\title{Stabilized Radiation Pressure Dominated Ion Acceleration from Thin-foil Targets}

\author{Min Chen\footnotemark[1], Naveen Kumar, Alexander Pukhov}
\affiliation{Institut f\"ur Theoretische Physik I,
Heinrich-Heine-Universit\"at D\"usseldorf, D\"usseldorf 40225,
Germany}

\begin{abstract}
We study transverse and longitudinal electron heating effects on the target stability and the ion spectra in the radiation pressure dominated regime  of ion acceleration by means of multi dimensional particle-in-cell~(PIC) simulations. Efficient ion acceleration occurs when the longitudinal electron temperature
 is kept as low as possible. However, tailoring of the transverse electron temperature is required in view of suppressing the transverse instability, which can keep the target structure intact  for longer duration during the acceleration stage. We suggest using the surface erosion of the target to increase the transverse temperature, which improves both the final peak energy
and the spectral quality of the ions in comparison  with a normal flat target.
\end{abstract}


\maketitle

In recent years, ion acceleration from thin foil targets has emerged as one of key areas of research in the field of laser plasma interaction~\cite{ion-acce}. The laser foil target interaction can produce energetic ions with energies as high as 56MeV per nucleon. Target engineering is crucial to get the mono-energetic ion beams~\cite{mono-ions}. These ion beams have extremely short duration ($\sim$fs), highly collimated and are relatively easy to produce, which makes  them suitable for many applications, such as proton imaging~\cite{ionimaging}, ion
therapy~\cite{therapy}, ion beam ignition of laser fusion
targets~\cite{fusion} and so on. It was recently suggested to utilize the circularly polarized (CP) laser pulses  for very high energy ion acceleration ~\cite{Macchi2005}. In the case of CP pulse, the electron heating is
dramatically reduced, thus ion acceleration doesn't occur due to the target normal sheath acceleration~(TNSA), instead
radiation pressure acceleration~(RPA) dominates. Though energy scalings predicted by 1D theory of RPA regime of acceleration have been reproduced very well in numerous one dimensional particle-in-cell~(PIC) simulations, the accelerated target is
far less stable in a real 3D geometry compare to 1D geometry. For instance, in the case of a Gaussian pulse interacting with a flat target, target
deformation occurs, which leads to the broadening of the final energy spectrum, and reduces the
energy conversion efficiency; though it can be overcome by the use of
shaped targets~\cite{Chen2009}. Another major bottleneck in the ion acceleration is the excitation of the Rayleigh-Taylor like instabilities, which leads to the breaking of the target~\cite{Pegoraro2007,Chen2008}. Thus, efficacy of the RPA scheme is limited by the onset of the Rayleigh-Taylor like instability; which also appears to be a cause of great concern in ion beam driven fast ignition. Prevention of this instability is an important problem which needs to be addressed quickly.

As described above, the RPA mechanism dominates when the electron heating is suppressed; this condition can be fairly met if one resorts to the use of the CP laser pulses. However, the onset of the Rayleigh-Taylor like instability, also known as transverse instability, remain a concern for the CP laser pulses. With regards to electron temperature, there is tradeoff as large perpendicular temperature might suppress the Rayleigh-Taylor like instability while longitudinal electron temperature can lead to the broadening of the energy spectrum. In this
paper, we explore such possibilities and study the heating effects by differentiating the
transverse and longitudinal heating of the electrons. We find that for
efficient ion acceleration, longitudinal heating of the electrons should be kept as low as
possible, however, proper transverse heating can suppress the
transverse instability and keep the target structure uniform for longer durations. Since the CP laser pulse can reduce
both transverse electron temperature $T_{e\perp}$ and longitudinal
electron temperature $T_{e\parallel}$, we suggest using target
surface erosion to increase $T_{e\parallel}$. We demonstrate, through multi-dimensional PIC  simulations, that this active
controllable method can keep the target structure uniform for
longer time compared with a normal flat target. We believe that our
differential treatment of the transverse and longitudinal electron
heating not only benefits the ion acceleration but could also be
useful in the fast ignition scheme.

\begin{figure}[t]
\begin{centering}\includegraphics[clip,width=80mm]{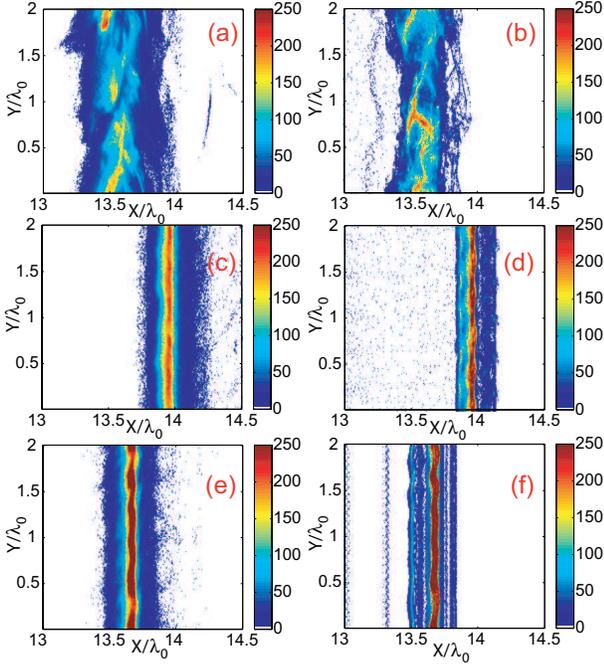} \par\end{centering}
\caption{(color online)\label{density_distribution} Spatial density
distribution of electrons~(left column) and protons~(right column)
at $t=20T_0$. (a,b) a flat target with initially zero temperature; (c,d) a modulation
target~($l_{depth}=0.05\lambda_0,\lambda_y=0.4\lambda_0$) with
initial zero temperature; (e,f) a flat target with initial transverse temperature of $T_e=6.3MeV$}
\end{figure}

We first show the results of 2D-PIC simulations performed by using the VLPL-code~\cite{Pukhov-vlpl}. The size of the simulation box is $15\lambda_0(x){\times}2\lambda_0(y)$ with
$\lambda_0$ representing the laser wavelength, which corresponds to
a grid of $3000(x){\times}200(y)$. The time step of the simulation
is $0.003T_0$, here $T_0=3.33$ fs is the laser period. The foil target consists of two species, electrons and protons. They are
initially located in the region $2\lambda_0\leq{x}\leq2.2\lambda_0$
with the density of $n=160n_c$, where $n_c=\omega^2m_e/4{\pi}e^2$ is
the critical density for the laser pulse with the frequency
$\omega$; which is  $n_c=1.1\times10^{21}/cm^3$ for 1${\mu}$m wavelength laser pulse. We use $150$ particles per cell to run the simulations. To exclude other effects arising due to the shape of the laser pulse, we here only take the plane
wave laser pulse. The normalized amplitude of the laser electric
field is $a_0=eE_0/m_e{\omega}c=100$. The laser pulse
has a trapezoidal temporal intensity profile (linear growth -
plateau - linear decrease), with $1\lambda_0/c - 8\lambda_0/c -
1\lambda_0/c$. At $t=0$ the laser pulse enters the simulation box
from the left boundary. Two kinds of targets are used: a normal flat target with initially zero temperature, and a target with surface erosion
and initially zero temperature. The second target has a ripple on the left surface, whose boundary is defined as $x(y)=x_0-l_{depth}{\times}[\sin(2{\pi}y/\lambda_y)+1]/2$ and $l_{{depth}}=0.05\lambda_0,\lambda_y=0.4\lambda_0$.

\begin{figure}[t]
\begin{centering}\includegraphics[clip,width=80mm]{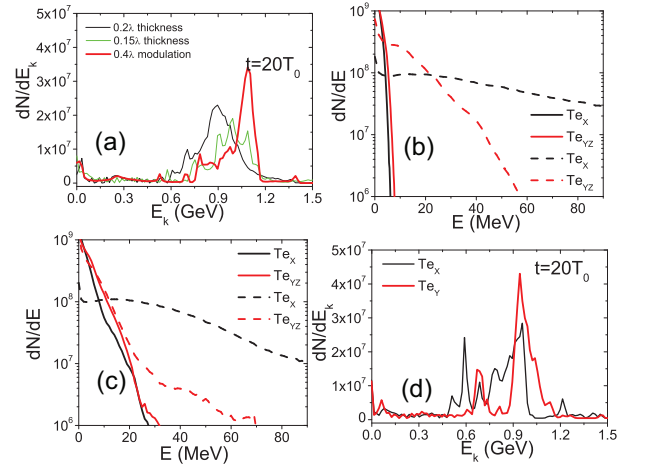} \par\end{centering}
\caption{(color online)\label{ele-ion-spec} (a) Proton spectrum
in the cases of a flat target with thickness of
$0.2\,\lambda_0$~(black solid line), $0.15\,\lambda_0$~(green solid
line) and a target with modulation~(red solid line) at $t=20\,T_0$. (b) Longitudinal ($T_{ex}$ , black line) and transverse ($T_{eyz}$, red/grey line) electron temperatures at $t = 5\,T_{0}$ (solid line) and $t = 20\, T_{0}$ (dashed line) for a flat target and initial zero temperature. (c) Longitudinal	($T_{ex}$ , black line) and transverse ($T_{eyz}$, red/grey line) electron temperatures at $t = 5\,T_{0}$ (solid line) and $t = 20\, T_{0}$ (dashed line) for a target with modulation of $\lambda_{y} = 0.4\lambda_{0}$ (black solid line) and initial zero temperature. (d)  Proton spectrum, at $t=20\,T_0$, in
the cases of initially longitudinal~(black solid line) and
transverse temperature~(red solid line). }
\end{figure}

Fig.~\ref{density_distribution} shows the spatial target density distribution
at $t=20T_0$. As we can see, the normal flat target  [subplots (a) and (b)] is almost completely dissociated after 20 laser periods. The target is
longitudinally expanded and transversely agglomerated. The flat target with an initial surface ripple [subplots (c) and (d)] has a better profile. Both the electrons and protons have sharp longitudinal boundaries and uniform transverse distribution. The main target has been longitudinally compressed to a layer of $0.07\lambda_0$ thickness. Its central position is $0.3\lambda_0$ forward than the normal flat target. The proton spectrums are shown in Fig.~\ref{ele-ion-spec}. The spectrum of the normal flat target is shown in Fig.~\ref{ele-ion-spec}(a) with the black solid line, while red line depicts the spectrum of the eroded target. It's clear that both the peak energy and width of the spectrum have been improved by using the surface erosion of the target. To exclude the target thickness effect, we also show the spectrum of a cold flat target with a thickness of $0.15\lambda$ with the green solid line. It also shows a broaden spectrum whose peak energy is smaller than the eroded target, which has a larger mass. Thus, it is clear that the eroded target can keep the acceleration structure intact for longer duration, thereby increasing the peak energy and reducing the width of the spectrum of the accelerated protons.

In order to ascertain the effect of electron temperature on the proton spectrum, we plot the target electron momentum distribution at different times for both targets. The electron temperature of the flat target and the surface eroded target at $t=5T_0$ (solid line) and at $t=20T_0$ (dashed line) are shown in Fig.~\ref{ele-ion-spec}(b) and (c), respectively. During the early stage of acceleration, the electrons in the flat target acquire lower longitudinal temperature ($T_{ex}$) as well as transverse temperature ($T_{eyz}$) compared with the surface eroded target. This implies that the target has been well compressed and the onset of instability has not occurred, resulting in target being pushed uniformly in the forward direction. The electrons in the surface eroded target attain higher temperatures perhaps due to higher absorption of the laser energy-caused by the surface ripples- into the target.   However, the longitudinal temperature is not high enough to break the target as the electrons' energies are far less than the ponderomotive potential level (~$a^2{\sim}$GeV), and most of the electrons are within the target area. However the scenario changes and at later times $20T_0$, we see both of the transverse and longitudinal electron temperatures are higher in the flat target than those of the surface eroded target. The higher electron temperatures in the flat target case are due to the breaking and subsequent heating of the target by the laser pulse. However, the transverse electron temperature of the surface eroded target hasn't witnessed higher growth. Most of the electrons in the target are around the proton layer to form the uniform high density plasma layer to suppress the laser penetration as also depicted in Fig. ~\ref{density_distribution} [subplot (c)]. One may infer during the early stage of acceleration ($T=5T_{0}$), the higher electron temperature in the surface eroded target appears to have provided a stabilizing influence on the transverse instabilities of the target, thus preventing the target from breaking.

In order to further isolate the temperature effects on the target structure during the acceleration stage, we perform another simulation by taking a flat target with an initial transverse temperature of 6.3MeV. The target electron and proton density distributions are shown in Fig.~\ref{density_distribution}(e) and (f). Again, we see a well accelerated target foil. The center of the target is little backward compared with the erosion target, which is due to the heavier mass of the target in this case. The proton spectrum at $t=20T_0$ is shown in Fig.~\ref{ele-ion-spec}(d) with the red solid line. It shows a narrow spectrum compared with the normal cold flat target case. We also check the effect of the longitudinal temperature by using a flat target with an initial longitudinal temperature of 6.3MeV. This temperature is not high enough to affect the RPA acceleration. However the final spectrum [see the black solid line in Fig.~\ref{ele-ion-spec}(d)] is even worse then the one in the normal flat target case. This means the longitudinal temperature cannot suppress the transverse instability and on the contrary it increases it.

From analytical view, to give a quantitative explanation one should use multi dimensional relativistic warm plasma theory with anisotropic relativistic temperature. For the moment, such kind of theory is still not available. Here we give a qualitative description by reducing the problem to two dimensional geometry and seeing the balance between the transverse temperature pressure and the perturbed light pressure. In the following we use normalized variables, such as: $x{\rightarrow}k_px, t{\rightarrow}\omega_pt,n{\rightarrow}n/n_c,p{\rightarrow}p/m_ec$. $k_p,\omega_p,n_c$ are the normal plasma wave number, frequency and critical density, respectively. We assume a circularly polarized laser pulse interacting with an initial flat target whose transverse relativistic temperature is $T_{ey}$. The density perturbation is in the transverse direction (in y direction) $n=n_0+{\delta}n$, and ${\delta}n=n_1\sin(2{\pi}y/{\lambda_y})f(x)$ is the second density perturbation term. The initial uniform electron density is $n_i$. For a short time ($t<<1/\omega_{pi}$), ions are assumed to be immobile. We will see the transverse temperature pressure can weaken the density perturbation due to light pressure. Because of two dimensional condition (x-y plane), we assume all the physical variables satisfy ${\partial}_z=0$. For a relativistic warm collision plasma, one can use Vlasov equation to derive the fluid equations. Here we use the fundamental results given by Schroeder \textit{et al.}\cite{Schroeder10} and use further assumption to simplify our model. Although our temperature is relativistic ($~6MeV$) we still adopt Schroeder's model to give a qualitative description. In further, in our case $\gamma_{th}<<\sqrt{1+a_0^2}$ (Here $\gamma_{th}$ corresponds to the temperature relativistic factor in Schroeder's paper and $a_0$ is the laser intensity we used here.), we neglect the thermal correction to the relativistic Lorentz factor as shown in formula (30) in Ref.\cite{Schroeder10}. The second term of relativistic temperature pressure in formula (30) has also been neglected due to time averaging over the laser cycle.

Then from the anisotropic warm plasma fluid equation, one can get the first and second order of the laser cycle averaged electron fluid equations as following.

The continuous equation gives:
\begin{eqnarray*}
  \frac{\partial{n_0}}{\partial{t}}+\bigtriangledown\cdot(n_0\vec{v_0}) &=& 0 \\
  \frac{\partial{\delta{n}}}{\partial{t}}+\frac{\partial{n_0}}{\partial{x}}\delta{v_x}+n_0\frac{\partial\delta{v_x}}{\partial{x}}+n_0\frac{\partial\delta{v_y}}{\partial{y}}+\frac{\partial\delta{n}}{\partial{y}}v_{0y}&=& 0
\end{eqnarray*}
We have already assumed $v_{0x}=0$ and $\partial{v_{0y}}/\partial{y}=0$, $\partial{n_{0}}/\partial{y}=0$ here, then we can get $\partial{n_0}/\partial{t}=0$, which shows the first order of balance between the light pressure and the static electric field resulting from the charge separation. However, the second order of density $\delta{n}$ is the density perturbation due to the perturbed light pressure.
The Poison equation gives:
\begin{eqnarray*}
  \nabla^2\phi_0 &=& n_0-n_i \\
  \nabla^2\delta\phi &=& \delta{n}
\end{eqnarray*}

For the motion of electron fluid, we have:
\begin{equation}\label{particle_motion}
    \frac{\partial\vec{p}}{\partial{t}}+\frac{\vec{p}}{\gamma}\cdot\nabla\vec{p} = \frac{\partial{\vec{a}}}{\partial{t}}+\nabla\phi-\frac{\vec{p}}{\gamma}\times\nabla\times\vec{a}+\frac{\nabla\cdot(\gamma^2\widetilde{n}<u_{ti}u_{tj}>)}{n}
\end{equation}
Here $u_{ti}$ is the normalized electrons' thermal momentum in the $\widehat{e}_i$ direction, the bracket means taking the average over its distribution, $\widetilde{n}$ is the proper density of the fluid element.
Considering the first order of motion (which generates the main electron density perturbation: $n_0-n_i$), since we have assumed $\partial{y}=0$, for the longitudinal direction we have:
$[\nabla\cdot(\gamma^2\widetilde{n}<u_{ti}u_{tj}>)]|_x={\partial_{\underline{x}}}(\gamma_0^2\widetilde{n}_0<u_{t\underline{x}}u_{tx}>)=<u_{tx}^2>\partial_x{(\gamma_0^2\widetilde{n}_0)}+\gamma_0^2\widetilde{n}_0\partial_x<u_{tx}^2>/2$, where the underline means the derivation only works on that variable. For the transverse direction we have $[\nabla\cdot(\gamma^2\widetilde{n}<u_{ti}u_{tj}>)]|_y={\partial_{x}}(\gamma_0^2\widetilde{n}_0<u_{tx}u_{ty}>)=0$, since $<u_{tx}u_{ty}>=<u_{tx}><u_{ty}>=0$ when the thermal momenta in different directions are not correlated.
For our interests, $<u_{tx}^2>{\ll}<u_{ty}^2>$ (equally to say $T_{ex}{\ll}T_{ey}$), thus we assume the longitudinal temperature effect could also be neglected, so in the first order of electron motion
there is no temperature effect.
The first order of electron motion equation is as following:
\begin{eqnarray*}
  \frac{\partial\vec{p}_0}{\partial{t}}+\frac{\vec{p}_0}{\gamma_0}\cdot\nabla\vec{p}_0 = \frac{\partial{\vec{a}_0}}{\partial{t}}+\nabla\phi_0-\frac{\vec{p}_0}{\gamma_0}\times\nabla\times\vec{a}_0
\end{eqnarray*}
In further we have:
\begin{eqnarray*}
  \frac{\partial(\vec{p}_0-\vec{a}_0)}{\partial{t}}=\nabla(\phi_0-\gamma_0)
\end{eqnarray*}

And $\vec{p}_{0\perp}=\vec{a}_{0\perp}$, $\vec{p}_{0x}=\vec{a}_{0x}=0$, $\gamma_0=\sqrt{1+a_0^2}$.

The second order of electron motion equation is:
\begin{eqnarray*}{\label{secondordermotion}}
  \frac{\partial(\delta\vec{p}-\delta\vec{a})}{\partial{t}}=\nabla(\delta\phi-\delta\gamma)+\frac{\vec{p}_0}{\gamma_0}\times\nabla\times(\delta\vec{p}-\delta\vec{a})\\+\frac{\nabla\cdot[(\gamma_0^2\delta{\widetilde{n}}+2\gamma_0\widetilde{n}_0\delta\gamma)<u_{ti}u_{tj}>]}{n_0}
\end{eqnarray*}

For the last term, again we have:
$\{\nabla\cdot[(\gamma_0^2\delta{\widetilde{n}}+2\gamma_0\widetilde{n}_0\delta\gamma)<u_{ti}u_{tj}>]\}|_x=\partial_{\underline{x}}[(\gamma_0^2\delta{\widetilde{n}}+2\gamma_0\widetilde{n}_0\delta\gamma)<u_{t\underline{x}}u_{tx}>]+\partial_{{y}}[(\gamma_0^2\delta{\widetilde{n}}+2\gamma_0\widetilde{n}_0\delta\gamma)<u_{t{y}}u_{tx}>]\simeq0$ since $<u_{tx}>\simeq0$ and $<u_{tx}^2>\simeq0$, and $\{\nabla\cdot[(\gamma_0^2\delta{\widetilde{n}}+2\gamma_0\widetilde{n}_0\delta\gamma)<u_{ti}u_{tj}>]\}|_y=\partial_x[(\gamma_0^2\delta{\widetilde{n}}+2\gamma_0\widetilde{n}_0\delta\gamma)<u_{tx}u_{ty}>]+<u_{ty}^2>\partial_y(\gamma_0^2\delta{\widetilde{n}}+2\gamma_0\widetilde{n}_0\delta\gamma)+(\gamma_0^2\delta{\widetilde{n}}+2\gamma_0\widetilde{n}_0\delta\gamma)\partial_y(<u_{ty}^2>)/2=<u_{ty}^2>[\gamma_0^2\partial_y(\delta{\widetilde{n}})+2\gamma_0{\widetilde{n}_0}\partial_y(\delta\gamma)]$ since $<u_{ty}>=0$ and $\partial_y(<u_{ty}^2>)=0$, and $\{\nabla\cdot[(\gamma_0^2\delta{\widetilde{n}}+2\gamma_0\widetilde{n}_0\delta\gamma)<u_{ti}u_{tj}>]\}|_z=0$ since $\partial_z=0$ and $<u_{tz}>=0$. The second term on the right side of Eq.~[\ref{secondordermotion}] can also be omitted by taking the laser cycle average for $\vec{p}_0/\gamma_0$.

Then one can get:
\begin{eqnarray*}\label{nebalance}
    \frac{\partial^2{\delta}n}{{\partial}t^2}+\frac{n_0}{\gamma_0}{\delta}n = \frac{n_0}{{\gamma_0}^2}\nabla^2(a_0{\delta}a)\\+{<u_{ty}^2>}[\gamma_0\frac{\partial^2\delta{n}}{\partial{y^2}}+2n_0\frac{\partial^2\delta\gamma}{\gamma_0{\partial}y^2}]+F(\delta{v_x},\delta{v_y})
\end{eqnarray*}
And the last term $F$ is not related to laser pressure and thermal pressure.
\begin{figure}[t]
\begin{centering}\includegraphics[clip,width=60mm]{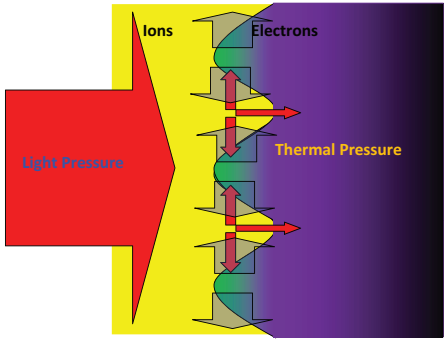} \par\end{centering}
\caption{(color online)\label{balance} Schematic profile of laser foil interaction. The red arrow shows the laser pressure and perturbed light pressure, the transparent blue arrow shows the thermal pressure.}
\end{figure}

Although the explicit relationship between $a{\delta}a$ and ${\delta}n$ has not been found yet, from Eq.~[\ref{nebalance}] one can see the thermal pressure can weaken the density modulation due to radiation pressure. In the region where the density is lower and the laser intensity is higher, the light pressure tends to expel the electrons in further, however, the thermal pressure tends to weaken such kind of plasma expelling. With a proper transverse temperature, the balance of thermal pressure and the perturbed light pressure can be built up just as shown in Fig.\ref{balance}. Thus the excitation of the instability can be effectively suppressed. Since the thermal pressure is used only to balance the perturbation of the laser pressure which is small, a transverse temperature of MeV level is enough. We do not need the GeV level of temperature to compete with the total laser pressure.

One may argue that the transverse temperature also causes transverse expansion of the target region. It seems the suppression of the instability would inevitably result into the target mass loss in the transverse direction. This is true. However, this kind of loss is far slower than the loss due to the instability. Moreover, recently Bulanov \textit{et al.} have argued that this kind of transverse loss could actually benefit the remaining target acceleration, and results into the accelerated ions getting phase locked with the electromagnetic wave. This could, in principle, produce unlimited ion energy gain~\cite{Bulanov10}. Certainly, the stable acceleration structure should be kept intact before the phase locking could actually occur.

In summary, we have demonstrated that the proper transverse temperature can suppress the transverse instability during the radiation pressure acceleration regime. We have shown that the surface erosion of the target can help in increasing the  transverse electron temperature. This scheme keeps the target intact for a longer duration  compared with a normal flat target, resulting into the improvement of both final peak energy and spectrum quality of the accelerated ions.

This work is supported by the DFG through TR-18 project. MC acknowledges
support by the Alexander von Humboldt Foundation and help from C.B. Schroeder on the relativistic warm plasma model.

\renewcommand{\thefootnote}{\fnsymbol{footnote}}
\footnotetext[1]{Present address: Lawrence Berkeley National
Laboratory, 1 Cyclotron Road, Berkeley CA, 94720. Email:
MinChen@lbl.gov}
%\bibliography{ionaccepaper}% Produces the bibliography via BibTeX.

\begin{thebibliography}{99}
\bibitem{ion-acce} T. Ditmire, \textit{et al.}, Nature \textbf{386}, 54
(1997); J. Denavit, Phys. Rev. Lett. \textbf{69}, 3052 (1992); A. Pukhov, \textit{ibid.} \textbf{86}, 3562 (2001); B.M. Hegelich, \textit{et al.},
\textit{ibid.} \textbf{89}, 085002 (2002); T. Esirkepov, \textit{et
al.}, \textit{ibid.} \textbf{89}, 175003 (2002); T. Esirkepov,
\textit{et al.}, \textit{ibid.} \textbf{92}, 175003 (2004); L. O. Silva,
\textit{et al.}, \textit{ibid.} \textbf{92}, 015002 (2004); J.
Fuchs, \textit{et al.}, Nature Phys. \textbf{2}, 48 (2006); Y. Lin, \textit{et al.}, Phys. Plasmas, \textbf{14}, 056706
(2007); M. Chen, \textit{et al.}, \textit{ibid.} \textbf{14}, 113106
(2007).
\bibitem{mono-ions}B.M. Hegelich, \textit{et al.}, Nature (London) \textbf{439}, 441
(2006);T. Toncian \textit{et al.} Science \textbf{312}, 410 (2006).
\bibitem{ionimaging}M. Borghesi, \textit{et al.}, Phys. Plasmas \textbf{9},
2214 (2002).
\bibitem{therapy}E. Fourkal, \textit{et al.}, Med. Phys. \textbf{34},
577 (2007).
\bibitem{fusion}N. Naumova, \textit{et al.}, Phys. Rev. Lett. \textbf{102}, 025002 (2009); M.
Roth \textit{et al.}, Phys. Rev. Lett. \textbf{86}, 436 (2001);
\bibitem{Macchi2005}A. Macchi, \textit{et al.}, Phys. Rev. Lett. \text{94}, 165003
(2005); X. Zhang, \textit{et al.}, Phys. Plasmas \textbf{14}, 123108
(2007); A. P. L. Robinson, \textit{et al.}, New J. Phys.
\textbf{10}, 013021 (2008); O. Klimo, \textit{et al.}, Phys. Rev.
Special Topics - Accelerators and Beams \textbf{11}, 031301 (2008);
X.Q. Yan, \textit{et al.}, Phys. Rev. Lett. \textbf{100}, 135003 (2008); A. Henig, \textit{et al.}, Phys. Rev. Lett. \textbf{103}, 245003 (2009);
A. Macchi, \textit{et al.}, New J. Phys. \textbf{12} 045103 (2010).
\bibitem{Chen2009} M. Chen, \textit{et al.}, Phys. Rev. Lett. \textbf{103},
024801 (2009); M. Chen, \textit{et al.}, New J. Phys. \textbf{12}, 045004 (2010).
\bibitem{Pegoraro2007}F. Pegoraro, \textit{et al.}, Phys. Rev. Lett. \textbf{99}, 065002
(2007).
\bibitem{Chen2008} M. Chen, \textit{et al.}, Phys.
Plasmas \textbf{15}, 113103 (2008).
\bibitem{Pukhov-vlpl}A. Pukhov, J. Plasma Phys. \textbf{61}, 425
(1999).
\bibitem{Schroeder10} C. B. Schroeder, E. Esarey, Phys. Rev. E \textbf{81}, 056403 (2010).
\bibitem{Bulanov10}S. V. Bulanov, E. Yu. Echkina, T. Zh. Esirkepov, I. N. Inovenkov, M. Kando, F. Pegoraro, and G. Korn, Phys. Rev. Lett. \textbf{104},
135003 (2010).
\end{thebibliography}

\end{document}